\documentclass[journal,comsoc,twocolumn]{IEEEtran}

\usepackage{booktabs} 
\usepackage[ruled,linesnumbered]{algorithm2e} 
\usepackage{multirow}
\usepackage{makecell}
\usepackage{tabulary}
\usepackage[normalem]{ulem}
\usepackage{amsmath,amsthm,amssymb}
\usepackage{mathtools}
\usepackage[caption=false]{subfig}
\usepackage{ulem}
\usepackage{pdfcomment}
\usepackage{enumitem}
\usepackage{textcomp}

\makeatletter
\newlength \figwidth
\if@twocolumn
\else
\fi
\makeatother

\newcommand\smsd{\bgroup\markoverwith{\textcolor{red}{\rule[0.4ex]{2pt}{0.8pt}}}\ULon}

\newcommand\dsd{\bgroup\markoverwith{\textcolor{blue}{\rule[0.4ex]{2pt}{0.8pt}}}\ULon}

\begin{document}
\title{Experimentally-based Cross-layer Optimization for Distributed Wireless Body-to-Body Networks}

\author{Samiya~M.~Shimly, 
        David~B.~Smith, 
        and Samaneh~Movassaghi 
        
\vspace{-0.5em}\thanks{Samiya M. Shimly, David B. Smith, and Samaneh Movassaghi are with the Research School of Engineering, The Australian National University (ANU), Canberra, ACT 0200, Australia. They are also with CSIRO (Commonwealth Scientific and Industrial Research Organisation) Data61, Australia. email: \{Samiya.Shimly, David.Smith, Samaneh.Movassaghi\}@data61.csiro.au}%
\thanks{This research is supported by an Australian Government Research Training Program (RTP) scholarship.}
}%

\maketitle

\begin{abstract}
We investigate cross-layer optimization to route information across distributed wireless body-to-body networks, based on real-life experimental measurements. At the network layer, the best possible route is selected according to channel state information (e.g., expected transmission count, hop count) from the physical layer. Two types of dynamic routing are applied: shortest path routing (SPR), and cooperative multi-path routing (CMR) associated with selection combining. An open-access experimental dataset incorporating `everyday' mixed-activities is used for analyzing and comparing the cross-layer optimization with different wireless sensor network protocols (i.e., ORPL, LOADng). Negligible packet error rate is achieved by applying CMR and SPR techniques with reasonably sensitive receivers. Moreover, at 10\% outage probability, CMR gains up to 8, 7, and 6 dB improvements over ORPL, SPR, and LOADng, respectively. We show that CMR achieves the highest throughput (packets/s) while providing acceptable amount of average end-to-end delay (47.5 ms), at -100 dBm receive sensitivity. The use of alternate paths in CMR reduces retransmissions and increases packet success rate, which significantly reduces the maximum amount of end-to-end delay and energy consumption for CMR with respect to other protocols. It is also shown that the combined channel gains across SPR and CMR are gamma and Rician distributed, correspondingly.
\end{abstract}%
%

%
%

\IEEEpeerreviewmaketitle

\section{Introduction}
Wireless body area networks (BANs) are often specifically designed for health-care scenarios to autonomously connect various medical sensors and actuators located on, in, around or/and near the human body to monitor physiological signals. Although, BAN applications span a wide area from military, ubiquitous health care, sports, entertainment to many more \cite{movassaghi2014wireless}, advanced professional health care management is one of the main purposes of the BAN concept. However, the underlying technology is still at an early stage of deployment and typically based on very specific wireless communications technologies \cite{filipe2015wireless}. The IEEE $802.15.6$ BAN Standard aims to enable low-power communication to be reliable and practical for in-body/on-body nodes to serve a variety of medical and non-medical applications \cite{tg6_std}. Now patients with BANs can be monitored while carrying out their regular everyday activities, without being tethered to monitoring devices.

With the anticipated growth in the number of people using BANs, their co-existence will be a concern in the near future, where reliable communications is vital in healthcare scenarios particularly. When multiple closely-located BANs coexist, the potential inter-network communication and cooperation across BANs lead to the investigation and design of wireless body-to-body networks (BBNs) \cite{meharouech2015future, meharouech2016two}. The main motivation behind BBNs is to make use of body-to-body (B$2$B) communications to overcome the problems of coexistence and general performance degradation for closely located BANs. As stated in \cite{meharouech2015future}, a body-to-body network is theoretically a mesh network that uses people (with body-worn sensors) to transmit or relay data within a limited geographic area, by creating their own centralized or decentralized network connection. This type of self-organizing, intelligent network could provide cost-effective solutions for remote monitoring of a group of patients wearing BAN sensors, for instance, by relaying physiological data/information up to the access point of the medical service, without depending on any external coordination. Some of the many target applications that can be benefited through BBN include: precision monitoring of athletes; facilitating rescue/medical teams in a disaster area; communication between groups of soldiers on a battlefield; and communication in densely populated areas such as city centers, concerts and sports venues. 

The notion of BBN is more dynamic and potentially larger-scale than that of individual BANs, where each BAN member can join and/or leave the network seamlessly, without the need for any centralized infrastructure. Hence, dynamic routing is necessary to enable routers to select paths according to real-time logical network layout changes  by periodic or on-demand exchange of routing information. Additionally, due to the postural body movements affecting BANs as well as their mobility, so that these BBNs are mobile, the use of networks incorporating infrastructure such as backbone routers would generally be very costly and unfeasible. Also, even in indoor BAN coexistence with mobile people, while deploying a backbone router is possible to some extent, the connection between a BAN hub and the router can be blocked by any obstacle or may go out of range because of the movement (e.g., postural changes or mobility). Specially, people having mobility issues can incur life-threatening risks, as these type of outages can be for a longer period of time with only a simple postural change. In such cases, a cooperative path through a nearby BAN hub, which is in the range of both source and destination hubs/backbone router, can greatly help to improve the situation. It is envisioned that in the case of unavailable or out-of-range network infrastructures, the BAN coordinators along with the BAN sensors can exploit cooperative and multi-hop body-to-body communications to extend the end-to-end network connectivity \cite{arbia2015behavior}.

\emph{In this paper, we seek to address the general performance of closely located BANs while utilizing body-to-body communications to extend end-to-end network connectivity across BANs without central coordination and interference mitigation. Hence, we perform and analyze energy efficient cross-layer optimization across the physical and network layers for two-tiered communications, with on-body BANs at the lower tier and a BBN at the upper tier to enable real-time, reliable human monitoring and communications across narrowband BANs}. The method is applied to an experimental radio measurement dataset\footnote{available in http://doi.org/10.4225/08/5947409d34552 \cite{smith2012body}} recorded from `everyday' mixed-activities and a range of measurement scenarios with people wearing radios. The extensive radio channel data was captured using NICTA\footnote{National Information and Communications Technology Australia (NICTA) has been incorporated into Data61, CSIRO.} developed wearable channel sounders/radios. We compare the results of cooperative multi-path routing (CMR) with shortest path routing (SPR) and other state-of-the-art WSN protocols (i.e., ORPL \cite{duquennoy2013let}, LOADng \cite{clausen2012loadng}). Our key findings in this paper, based on empirical results derived from real-life measurements \cite{smith2012body}, are as follows:
\begin{itemize}
    \item[$\bullet$] Negligible (almost $0$\%) packet error rate (less than $10$\%, thus fulfilling the requirement of the IEEE $802.15.6$ Standard \cite{tg6_std}) is achieved with reasonably sensitive receivers for both shortest path routing (SPR) and cooperative multi-path routing (CMR) in a dynamic environment associated with mobile subjects, using the available nodes/hubs as relays.
     \item[$\bullet$] CMR provides up to $8$ dB, $7$ dB, and $6$ dB performance improvement over ORPL\footnote{We have implemented these protocols (i.e., ORPL, LOADng) in MATLAB and applied on the same measurement dataset (used in this paper) to compare with SPR and CMR.\label{footnote 3}}, SPR, and LOADng\textsuperscript{\ref{footnote 3}}, respectively, at $10$\% outage probability.
     \item[$\bullet$] ETX (Expected Transmission Count) or hop count metric (used in SPR, CMR, and LOADng in a mesh) can perform better than EDC (Expected Duty Cycles) metric (used in ORPL with DODAG\footnote{Destination Oriented Directed Acyclic Graph \cite{gnawali2009collection,buonadonna2004multihoplqi}.} topology) in case of any-to-any routing among BANs.
     \item[$\bullet$] CMR outperforms other protocols in case of throughput (packets/second) by providing $95\%$ successful packet delivery ($19$ packets/s at a packet transmission rate of $20$ Hz).
    \item[$\bullet$] The maximum amount of end-to-end delay is the lowest for CMR ($135$ ms) with respect to other protocols and also below the IEEE $802.15.6$ latency requirement (< $250$ ms) for non-medical applications. Also, the average end-to-end delay for CMR ($47.5$ ms) is an acceptable amount (< $125$ ms) for BAN medical applications.
    \item[$\bullet$] CMR consumes more energy on average than other techniques due to the cooperative combining at route-hops, although the maximum energy consumption with CMR is much lower than other protocols (except SPR with hop restriction).
    \item[$\bullet$] CMR produces the lowest amount of average end-to-end delay with respect to other protocols, when estimated with lower receive sensitivity (e.g., $-90$ dBm, $-86$ dBm).
    \item[$\bullet$] With less receive sensitivity (e.g., $-90$ dBm, $-86$ dBm), the energy consumption of CMR remains relatively similar, while the energy consumption of other protocols increases significantly due to an increase in packet failure rate and retransmissions.
    \item[$\bullet$] The empirical received signal amplitude through SPR has a gamma distribution while the empirical received signal amplitude through CMR has a Rician distribution.
\end{itemize}
The rest of this paper is organized as follows: In Section $2$, work related to cross-layer optimization for wireless body area networks is described. The system model with the experimental scenario is presented in Section $3$ along with the description of the routing techniques incorporated with physical layer processing. The experimental results (i.e., outage probability, throughput, end-to-end delay, energy consumption per packet, percentage of hop count) for measuring the performance of the protocols are analyzed and compared with other protocols in Section $4$. Additionally, the probability distribution functions, fitted to combined channel gains after routing is applied are provided in this section. Section $5$ provides some concluding remarks.%

\section{Related Work}
Many studies have been focused on BANs for medical purposes \cite{jovanov2005wireless, fang2009bodymac, timmons2009adaptive, li2010heartbeat, tauqir2013non, rushanan2014sok, shimly2016cooperative}. As stated in \cite{filipe2015wireless}, few works have been concerned with a global solution for tens or hundreds of patients, each of whom is fitted with multiple sensor nodes, and confined to a relatively small environment. As literature has shown in \cite{movassaghi2013review}, even though the general characteristics of BANs are somewhat similar to mobile ad hoc networks (MANETs) \cite{abolhasan2004review} and wireless sensor networks (WSNs) \cite{akkaya2005survey}, the stringent requirements of BANs impose certain constraints on the design of their routing protocol which leads to novel challenges, which can not be met through typical WSN/MANET routing protocols. For example, the frequent topological changes in highly mobile coexisting BANs particularly occur with group-based movement, rather than node-based movement in MANETs, which suggests that all nodes in BANs move with keeping their position with respect to one another, while in MANET each node moves independently from other nodes in the network \cite{asgari2015overview}. In fact, the on-body sensor nodes used in BANs move relative to the coordinator node of the corresponding BAN which is used as their reference point \cite{movassaghi2016enabling}. As the user moves, the whole network moves and then may move into (and out of) the range of other networks frequently, which results in network collision \cite{cheng2009network}. It is different to the interference events of cellular/sensor networks where only one or two nodes interfere, and base stations rarely interfere \cite{hanlen2009interference}. The random nature of BAN movement means that network collisions can be very short (e.g., people passing on the street) or very long (e.g., family members/hospital patients may remain close for hours) \cite{cheng2009network}. Additionally, routing protocols specifically designed for MANETs are based on completely decentralized networks, whereas BANs can possess coordination at times, to lessen the overhead when possible. Moreover, BANs have more strict energy constraints in terms of transmit power compared to traditional sensor and ad hoc networks as node replacements can be quite uncomfortable and might require surgery in some scenarios (e.g. implant nodes) \cite{movassaghi2014wireless}.

In the past decade, several cluster-based and cross-layer routing protocols have been proposed for BANs along with other routing protocols \cite{movassaghi2013review}. Some of the cluster-based routing protocols (e.g. ANYBODY \cite{watteyne2007anybody}, HIT \cite{culpepper2003hybrid}) that have been designed for BANs aim to minimize the number of direct transmissions from sensors to the base station. WASP \cite{braem2006wireless}, CICADA \cite{braem2008improving}, TICOSS \cite{ruzzelli2007energy} and BIOCOMM \cite{bag2009biocomm} are some cross-layer protocols between Network and MAC layer for BANs. Amongst these protocols, TICOSS and CICADA consume less energy, whereas the WASP scheme outperforms others in terms of efficient packet delivery ratio (PDR) \cite{comparative2015}. Also, CICADA performs well among the other protocols in terms of reducing packet delivery delay \cite{comparative2015}. Otal et al. proposed an energy-saving MAC protocol, DQBAN (Distributed Queuing Body Area Network) for BANs in \cite{otal2009highly}, as an alternative to the 802.15.4 MAC protocol which suffers from low scalability, low reliability and limited QoS in real-time environments. The proposed DQBAN is a combination of a cross-layer fuzzy-logic scheduler and energy-aware radio-activation policies \cite{crosby2012wireless}. The fuzzy-logic scheduling algorithm is shown to optimize QoS and energy-consumption by considering cross-layer parameters such as residual battery lifetime, physical layer quality and system wait time \cite{crosby2012wireless}. A number of interference-aware coexistence schemes for multiple BANs have been proposed in \cite{dong2013opportunistic,movassaghi2016enabling}. In \cite{dong2013opportunistic}, a cooperative two-hop communication scheme together with opportunistic relaying (OR) is applied on a single BAN (in case of BAN coexistence), which improves the outage probability and level crossing rate of on-body channels with respect to suitable SINR threshold values. The authors in \cite{movassaghi2016enabling} proposed an energy efficient and interference-aware channel allocation scheme that also incorporates an intra-BAN and inter-BAN mobility model for BAN coexistence. A Cross-layer Opportunistic MAC/Routing protocol (COMR) \cite{abbasicross} has also been proposed for improving reliability in BAN, where the authors have used a timer-based approach with combined metrics of residual energy and receive signal strength indicator (RSSI) as their relay selection mechanism in a single BAN and compared it with Simple Opportunistic Routing (SOR) \cite{mao2011energy}. In \cite{tseng2016efficient}, the authors have proposed an efficient cross-layer reliable retransmission scheme (CL-RRS) without additional control overheads between physical (PHY) and MAC layer, which significantly improves frame loss rate and average transmission time as well as reduces power consumption.

Some WSN routing protocols for low power lossy networks (LLNs) are proposed in literature. For example, RPL \cite{winter2012rpl} uses a DODAG/rooted topology based on expected transmission count (ETX) metric which is similar to the collection tree protocol (CTP) proposed in \cite{gnawali2009collection} for WSNs, where the sink node collects data from different sensors with datapath validation and adaptive beaconing. In RPL, any-to-any routing is performed with a non-storing mode through the root node. To improve the performance of RPL, opportunistic routing protocol (ORPL) is proposed in \cite{duquennoy2013let}, which combines opportunistic routing with rooted topology, where any-to-any routing is supported through the common ancestors along with the root node based on EDC (Expected Duty Cycle) metric. In \cite{clausen2012loadng}, the authors proposed a reactive distance-vector routing protocol named LOADng, which inherits the basic properties and operation of AODV (Adhoc On-demand Distance Vector routing) \cite{perkins2003ad}, yet aims to reduce the per packet overhead for route discovery by smart route request \cite{yi2012smart} and expanding ring search \cite{bas2012expanding}. Cooperative multi-path routing \cite{liu2008cooperative} yields better performance than single-path routing by providing simultaneous parallel transmissions with load balancing over available resources. We proposed a new CMR scheme in \cite{shimly2017cross} for coordinated BANs, that uses two different paths (incorporating shortest path routing) which are combined at the destination. Also, it uses TDMA with low duty cycling to save energy consumption and avoid interference from surrounding non-coordinated BANs. However, reducing duty cycle or active period often increases the overall delay and applying scheduling techniques without global coordination incurs extra overheads. In this paper, we investigate the performance of CMR \cite{shimly2017cross} without any multiple access scheme or interference mitigation, by performing cross-layer optimization between PHY and Network layers for distributed BBN (or co-located BANs).

Most previous works have not considered practical BAN coexistence with any-to-any routing (without any specific root/coordinator), using actual measured data, for intra-BAN and inter-BAN (BBN) communications in tiered architecture. Additionally, previous routing protocols are mostly based on cross-layer optimization between adjacent layers (e.g., PHY-MAC-Network layers, PHY-MAC layers, MAC-Network layers) without violating the reference architecture (e.g., OSI model), hence wasting resources which can be optimized by non-adjacent cross-layer approaches. The non-adjacent cross-layer (PHY-Network layers) methods described in this paper can incorporate both postural body movements (intra-BAN communications) and mobility (inter-BAN communications) together with acceptable delay and energy efficient routing, including excellent communication reliability across BANs.%

\section{System model}
\begin{figure}[!t]
\centering{\includegraphics [width=\figwidth]{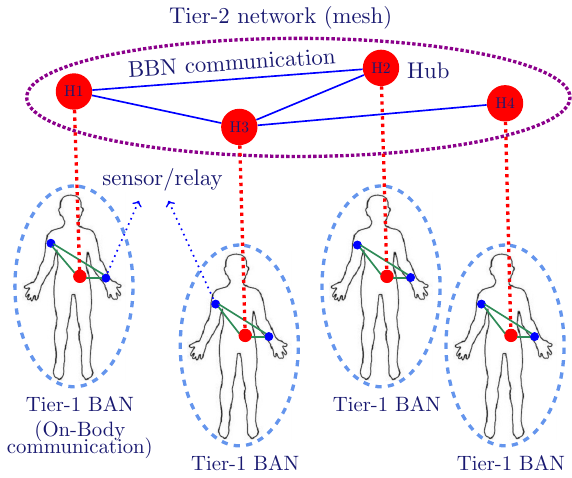}}
\caption{Two tiered architecture with 4 coexisting BANs; Hub on the left-hip
and two sensors/relays on the left-wrist and right-upper-arm, respectively.}
\label{Tier}
\end{figure}

\begin{figure*}[!htb]
\centering{\includegraphics [width=5in]{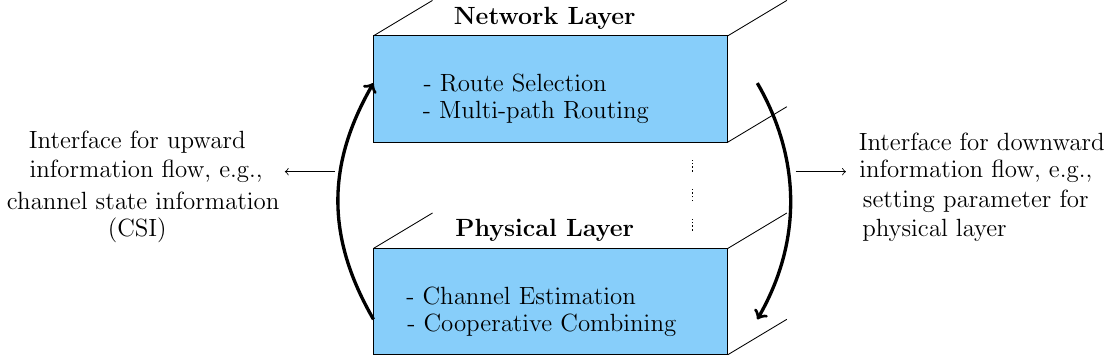}}
\caption{Cross-layer optimization between Physical and Network layers.}
\label{crosslayer}
\end{figure*}

We assume a two-tiered network architecture formed from $10$ co-located mobile BANs (people with fitted wearable radios) deployed for experimental measurements, where the hubs of the BANs are in tier-$2$ in a mesh (inter-BAN/ BBN communications) and the on-body sensors of the corresponding BANs are in tier-$1$ (intra-BAN communications). An abstraction of the architecture is given in Fig. \ref{Tier}, with four co-located BANs. 
It can be portrayed as a hybrid mesh architecture where BANs (hubs/gateways) are performing as both clients and routers/relays, which will enable flexible and fast deployment of BANs to provide greater radio coverage, scalability and mobility. When a node/BAN hub is unable to send information directly to the intended destination node/hub, it tries to relay the information through the nearby BAN hubs. A given node, when acting as relay, follows the decode-and-forward relaying scheme for which it decodes the signal and then retransmits it. Any-to-any routing is performed in a cross-layered approach, with two different routing techniques, i.e., shortest path routing (SPR) and cooperative multi-path routing (CMR), that utilize and interact with the physical layer. Therefore, changes in channel states (i.e., expected transmission counts, hop counts) are directly indicated from the physical layer to the network layer, so that the routes with most favorable channel conditions are chosen (illustrated in Fig. \ref{crosslayer}).
The cross-layer approach across the PHY--Network layer can be implemented by creation of new interfaces for upward/downward information flow \cite{srivastava2005cross}. For real-time dynamic estimation, we timestamp\footnote{The whole channel is divided into timestamps of a given period and the routing table updates after each timestamp. The samples are taken periodically over the timestamp period.} the samples of any given link continuously with a reasonable timestamp period of $500$ ms, given the longer coherence times of $500$ ms (up to $1$ s) for `everyday' activities of narrowband on-body BAN channels \cite{smith2013propagation}, and $900$ ms (calculated in \cite{sshimly2018cross}) for body-to-body channels used here. Therefore, the routing table for a given timestamp is updated based on the estimated channel condition from the previous timestamp. The experimental scenario and the routing techniques are described in the following subsections.%

\subsection{Experimental Scenario}
The open-access dataset \cite{smith2012body} stated in this paper consists of continuous extensive intra-BAN (on-body) and inter-BAN (body-to-body) channel gain data incorporating varying amount of movements for many tens of hours of measurements. A range of different carrier frequencies (near $400$, $900$ and $2400$ MHz) and communication bandwidths are used \cite{smith2011first}. We analyze the performance of the protocols upon a suitable portion of the dataset (can be downloaded from \cite{smith2012body}) of around $45$ minutes, captured from $10$ closely located mobile subjects (adult male and female). We consider $10$ mobile people (with intra-body and inter-body interactions) as a reasonable amount for BAN coexistence experimentation, due to the dramatic impact caused by the slowly-varying human-body dynamics and shadowing by body-parts, both on the on-body and inter-body channels \cite{smith2015channel,hanlen2010co,smith2008statistical}. Also this is an acceptable range of BANs to be supported by the physical layer according to the IEEE $802.15.6$ Standard \cite{tg6_std}. 

\begin{figure}[!t]
\centering{\includegraphics [width=2.5in]{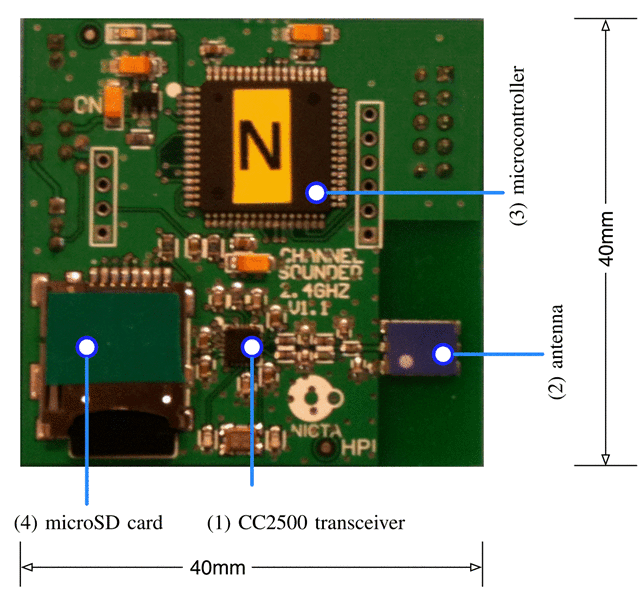}}
\caption{The radio-frequency testbed with major components highlighted. Battery (disconnected) is on reverse side.}
\label{Sen}
\end{figure}

The experimented subjects were walking together to a hotel bar, sitting there for a while and then walking back to the office. Each subject wore $1$ transmitter on the left-hip and $2$ receivers on the left-wrist and  right-upper-arm, respectively (shown in Fig. \ref{Tier}). The radios were transmitting at $0$ dBm transmit power with $-100$ dBm receive sensitivity. A detailed description of these wearable radios and hardware platform can be found in \cite{hanlen2010open}. As described in \cite{hanlen2010open}, the testbed used in the measurement (shown in Fig. \ref{Sen}) is a programmable radio transceiver with non-volatile data storage, designed to be worn by test subjects for several hours. The testbed is comprised of -- radio transceiver (Texas Instruments CC$2500$) which communicates digital data in the form of packets with the micro controller, Bluetooth surface-mount ceramic multilayer `chip' antenna (YAGEO CAN$4311111002451$K), micro controller (Atmel ATmega1281), microSD card socket and rechargeable battery (Energizer CP18NM). The microSD card is used to store gain measurements, which is transferred to a computer using an off-the-shelf card reader. Each transmitter sends packets with a transmit ID attached. When a channel sounder device receives a packet from another device it logs the packet ID and RSSI (receive signal strength indicator): the ID provides the link identification and the RSSI  provides the instantaneous signal strength for that link \cite{hanlen2009interference}. We use MATLAB for analyzing and investigating the performance of coexisting BANs with the experimental dataset. The parameters applied for estimating the performance metrics are listed in Table \ref{table_param}.
Each transmitter transmits in a round-robin fashion, at $2.36$ GHz (near ISM band, avoiding ISM interference), with $5$ ms separation between each other. Hence, each transmitter is transmitting in every $50$ ms (with a sampling rate of $20$ Hz) to every $9$ other subject\textquotesingle s receivers as well as their own receivers (all small body-worn radios/hubs/sensors), along with capturing the RSSI values in dBm, which gives a total of $300$ channel measurements (including both on-body and body-to-body links) in real-time over the whole network during the measurement period. Due to the reciprocity property, the channel from any $T_x$ (transmitter) at position $a$ to $R_x$ (receiver) at position $b$, is similar for $T_x$ at $b$ to $R_x$ at $a$ \cite{hanlen2010open}, thus transmitters and receivers can be considered interchangeably to model multiple synchronous networks \cite{hanlen2009interference}.
\begin{table}[!t]
\caption{Applied Parameters}
\label{table_param}
\begin{minipage}{\columnwidth}
\begin{center}
\begin{tabular}{|l|l|}\Xhline{1pt}
  Carrier Frequency & 2.36 GHz\\\Xhline{0.8pt}
  Data rate         & 486 kbps\\\hline
  Packet size       & 273 bits\\\hline
  Sampling rate     & 20 Hz\\\hline
  Packet transmission time ($T_{packet}$)    & 0.6 ms\\\hline
  Transmission period ($T_{active}$)        & 5 ms\\\hline
  Sampling period        & 50 ms\\\hline
  \begin{tabular}{@{}l@{}l@{}}Tx/Rx mode power consumption:\\on-body hub
  ($P_{TX_h}$/$P_{RX_h}$)\end{tabular}      & 6 mW\\\hline
  \begin{tabular}{@{}l@{}l@{}}Tx/Rx mode power consumption:\\on-body sensor
  ($P_{TX_s}$/$P_{RX_s}$)\end{tabular}     & 5 mW\\\hline
  Idle mode power consumption ($P_{idle}$) & 1 mW\\\hline
  Transmit power    & 0 dBm\\\hline
  Timestamp period  & 500 ms\\\hline
  Total time & 45 mins\\\Xhline{1pt}
\end{tabular}
\end{center}
\bigskip\centering
\end{minipage}
\end{table}%

\subsection{Shortest Path Routing (SPR)}
We perform dynamic shortest path routing (SPR) \cite{wang1992analysis} based on an Open Shortest Path First (OSPF) protocol, which uses link-state algorithm (e.g., Dijkstra's algorithm), where any source node intends to find route with a minimum cost (based on routing metrics) to any destination and updates the routing table dynamically to adapt variable channel conditions and topological changes. Routing metric predicts the cost of the route calculated from the use of a certain routing protocol, which plays a critical role in finding out the efficient route to the destination in a network. For investigating the performance of the dynamic routing, we use a combination of different routing metrics (e.g. ETX, hop count etc.). Although some well-known WSN protocols (e.g., CTP \cite{gnawali2009collection}, RPL \cite{winter2012rpl}) make use of the ETX metric, the calculation of ETX in our work differs with respect to the gradient descent technique followed by those protocols to calculate ETX. As these protocols (i.e., CTP, RPL) use a rooted topology (for collecting information), each node maintains an estimation of its route cost to a collection point/specific sink node, where the collection point exhibits a cost of zero, which leads to a convergence towards the root node, as a node only selects forwarding nodes that provide strictly more progress than itself with a lower ETX. By contrast, in our work, we simply find routes between two nodes (from any source to any destination) which have minimum cost (sum of the ETX values of the links in a route), without any back propagation to a root node.
\begin{figure}[!t]
\centering{\includegraphics [width=2.5in]{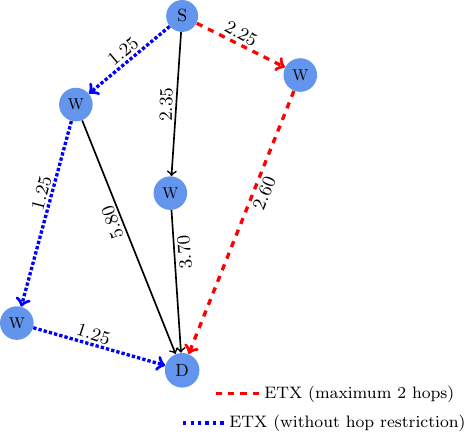}}
\caption{Shortest path routing (SPR), with and without hop restriction.}
\label{imagespr}
\end{figure}%

\begin{algorithm}[!t]
\SetKwFunction{length}{length}
\SetKwFunction{minimum}{minimum}
\SetKwFunction{PathHopCount}{PathHopCount}
\SetKwFunction{Path}{Path}
\SetKwFunction{isempty}{isempty}
\SetKwInOut{Input}{input}
\SetKwInOut{Output}{output}
\Indmm
\Input{Source node ($S$), destination node ($D$), and set of intermediate nodes ($N$).}
\Output{Shortest path from $S$ to $D$.}
\Indpp
\BlankLine
$P_{etx} \leftarrow$ Set of ETX values for every possible paths from $S$ to $D$\;
[$i,j$] $\leftarrow$ [$1$,\length{$P_{etx}$}]\;
\While{$i\ne j$}
{
    $etx_{min} \leftarrow$ \minimum{$P_{etx}$}\;

    \eIf{\PathHopCount{$etx_{min}$} = $2$}
    {
    	$Output_{SPR} \leftarrow$ \Path{$etx_{min}$}\;
    }
    {
        $P_{etx} \leftarrow P_{etx} - etx_{min}$\;
        $etx_{min} \leftarrow$ \minimum{$P_{etx}$}\;
        $j \leftarrow j - 1$\;
    }
    $i \leftarrow i + 1$\;
}
\If {\isempty{$Output_{SPR}$}}
{
	$Output_{SPR} \leftarrow$ direct path\;
}
\caption{Finding shortest path route (with ETX + maximum two hops count)}
\label{shortestpath}
\end{algorithm}
Furthermore, in RPL, if a source node wants to communicate with any node (other than root or sink node), the route has to go first upwards (towards the root node) and then downwards (towards the destination), as RPL uses only one common ancestor node instead of a full mesh (similar as CTP) \cite{duquennoy2013let}. This increases the number of hops in a route as well as the delay and energy consumption. In our work, we have restricted the routes to a maximum of two hops, which saves energy and latency while providing acceptable packet delivery ratio (demonstrated in following sections).
\paragraph{ETX} The Expected Transmission Count (ETX) path metric is a simple, proven routing path metric that favors high capacity and reliable links. This metric estimates the number of retransmissions required to send unicast packets by measuring the loss rate of broadcasted packets between pairs of neighboring nodes \cite{malnar2009comparison}, which can be calculated as follows:
\begin{equation}\label{etx}
ETX = \frac{1}{1-O_p},
\end{equation}
where $O_p$ is the outage probability with respect to the receiver sensitivity (e.g., $-100$ dBm). ETX adds more reasonable behavior under real life conditions, since this metric is based on packet loss and thus the number of packets sent.
\paragraph{Hop count} Hop count identifies the route which has minimal number of hops. The primary advantage of this metric is its simplicity. Once the network topology is known, it is easy to compute and minimize the hop count between a source and its destination. However, the primary disadvantage of this metric is that it doesn't take packet loss, bandwidth, power consumption or any other characteristic of a link into account \cite{malnar2009comparison}.

In this paper, an optimal path is selected by combining these two metrics (ETX + Hop count), restricting the hop count to two hops. A pseudo-code for this process is given in Algorithm \ref{shortestpath}. The ETX  of a given link is estimated from EQ. (\ref{etx}) with the RSSI values of the previous timestamp ($500$ ms) to be applied in the next timestamp of $500$ ms. As the router performs periodic updates, when there is no route from source to destination (ETX = $\infty$), the router chooses the direct link for the given timestamp period to possibly avoid longer delay. In SPR, the combined channel gains at the destination can be measured as follows:
\begin{equation}\label{spr_eqn}
spr_{comb} = min\big(h_1, ..., h_k\big),
\end{equation}%
where, $k$ is the number of hops in the route; $k = 2$ when two hop restriction is applied on SPR.
In Fig. \ref{imagespr}, an illustration of shortest path routing (SPR) is presented, where the difference between using the ETX metric with and without hop restriction is demonstrated. The path taken without hop restriction can be a longer path with the lowest cost. However, the path associated with combined metric (ETX + maximum two hops) is an optimal path, having the lowest cost possible with a maximum of two hops.%

\subsection{Cooperative Multi-path Routing (CMR)}
\begin{figure}[!t]
\centering{\includegraphics [width=2.5in]{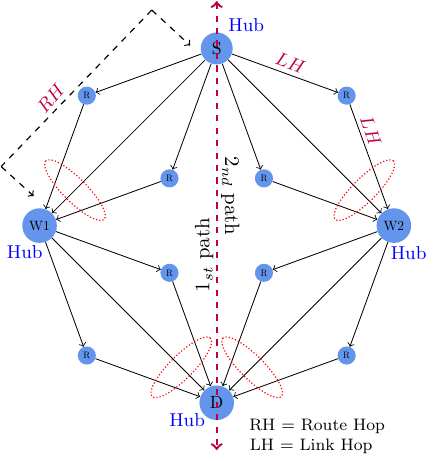}}
\caption{Cooperative multi-path routing (CMR) with $3$-branch selection combining in each route hop.}
\label{imagecmr}
\end{figure}%

\begin{algorithm*}[!t]
\SetKwFunction{FindShortestPathWithHopRestriction}{FindShortestPathWithHopRestriction}
\SetKwFunction{isempty}{isempty}
\SetKwFunction{SelectionCombining}{SelectionCombining}
\SetKwFunction{length}{length}
\SetKwFunction{minimum}{minimum}
\SetKwInOut{Input}{input}
\SetKwInOut{Output}{output}
\Indmm
\Input{Source node ($S$), destination node ($D$), and set of intermediate nodes ($N$).}
\Output{Result of CMR at destination node.}
\Indpp
\BlankLine
$P1 \leftarrow$ \FindShortestPathWithHopRestriction{$S, D, N$}\;
$P2 \leftarrow$ \FindShortestPathWithHopRestriction{$S, D, N$} $\neq P1$\;
\uIf {\isempty{$P2$} $\neq$ 1}
{
  \For {$i\leftarrow 1$ \KwTo $2$}
  {
    \For {$j\leftarrow 1$ \KwTo \length{$P_i$}}
    {
    	$Pi_{RH_j} \leftarrow$ \SelectionCombining{Route\_Hop$_j$}\;
    }
  }
  \uIf {\length{$P_1$} > $2$}
  {
  	$Comb_{P1} \leftarrow$ \minimum{$P1_{RH_1}, P1_{RH_2}$}\;
  }
  \Else
  {
  	$Comb_{P1} \leftarrow P1_{RH_1}$\;
  }
  $Comb_{P2} \leftarrow$ Repeat steps $9$ to $13$ for $P2$\;
  \uIf {$Comb_{P1}$ is successful}
  {
  	$Output_{CMR} \leftarrow Comb_{P1}$\;
  }
  \Else
  {
  	$Output_{CMR} \leftarrow Comb_{P2}$\;
  }
}
\Else
{
	Repeat steps $4$ to $14$\;
    \uIf {$Comb_{P1}$ is successful}
  	{
  		$Output_{CMR} \leftarrow Comb_{P1}$\;
  	}
  	\Else
  	{
  		Retransmit through P1\;
  	}
}
\caption{Estimating Output of CMR}
\label{OP_CMR}
\end{algorithm*}
In dynamic cooperative multi-path routing \cite{shimly2017cross}, we use cooperative paths from source to destination with combined channels in each route-hop. In this paper, route-hop refers to each hop of a path/route in CMR from source hub to destination hub through an intermediate BAN hub (acting as a mesh router/relay). In each route-hop, $3$-branch cooperative selection combining is used, where one of the branches is the direct link and the other two branches (cooperative relayed links) have two link-hops. Link-hop refers to each hop of the branch from a BAN hub through on-body relays of the corresponding BAN. A decode-and-forward protocol is applied at each on-body relay. The equivalent channel gain at the output of selection combining can be estimated as follows:
\begin{equation}\label{sc}
h_{sc}(\tau) = \max \Big\{{h_{sd}(\tau), h_{sr_1d}(\tau), h_{sr_2d}(\tau)}\Big\}
\end{equation}%
where $h_{sc}(\tau)$ is the equivalent channel gain at the output of the selection combining, at time instant $\tau$. $h_{sd}$ is the channel gain from source-to-destination (direct link), $h_{sr_id}= \min \{h_{sr_i},h_{r_i d}\}$ are the channel gains of the first and second cooperative relayed links (with two link-hops, $s$ to $r_i$, and $r_i$ to $d$), $i=[1,2]$, respectively.

For multi-path routing here, two different paths are used (if more than one path is available) from source hub to destination hub, where both paths can have maximum of two route-hops. The shortest path is chosen according to an SPR calculation, hence the two paths go through the two nearest BAN hubs from the source. The nearest BANs from any given source can be found from the source hub-to-connected hub channel gains, approximated from the RSSI at the connected BAN hubs. In CMR, the combined channel gains at the destination for a given path/route is measured as follows:
\begin{equation}\label{cmr_eqn2}
 cmr_{comb} = min\big\{h_{RH_1}, h_{RH_2}\big\}
\end{equation}
where $h_{RH_j}$ are the cooperatively combined (with selection combining) channel gains at route hop $j=[1,2]$ of the route. $h_{RH_j}$ can be calculated from Eq. (\ref{sc}).

In this paper, we apply a slightly different approach from that in \cite{shimly2017cross} at the destination of CMR, where the destination does not wait for a specific period for combining two different copies of the packets received from two different paths (if more than one path is available), instead the destination process the data/information once it receives the packet/data (through any of the paths). Also, for body-centric channels (specially for B$2$B channels), retransmissions of failed packets are often impractical (except for critical applications) \cite{boulis2012challenges} due to the longer coherence times. For instance, retransmitting a failed packet multiple times (with a retransmission timeout period) will increase the delay and energy consumption, with lower probability of successful delivery as the channel condition remains similar for hundreds of milliseconds. One of the main objective of cooperative multi-path routing is to reduce such retransmissions in order to avoid unnecessary delay. Here, retransmission is only permitted in CMR when the shortest path has a single route-hop and no alternate path is possible. However, in CMR, the use of an alternate path and the cooperative combining in each route-hop of a path increases the chance of successful packet delivery despite avoiding retransmissions (as shown in the following section). The process for CMR is illustrated in Fig. \ref{imagecmr} and described with a pseudocode in Algorithm \ref{OP_CMR}.%

\section{Experimentally-based Results}
In this section, we analyze the performance of the cross-layer approach and compare the results with different WSN protocols, i.e., ORPL \cite{duquennoy2013let}, LOADng \cite{clausen2012loadng} for low-power and lossy networks (LLNs) that support any-to-any routing. We implement ORPL and LOADng in MATLAB, and applied those protocols on the same measurement dataset \cite{smith2012body} for a fair comparison with SPR and CMR. When implementing ORPL and LOADng, two nodes are considered as neighbors if the link RSSI is greater than or equal to the receiver sensitivity (i.e., $-100$ dBm).%

\paragraph{ORPL} ORPL uses a combination of a rooted/DODAG (Destination Oriented Directed Acyclic Graph) topology with opportunistic routing based on the EDC (Expected Duty Cycles) metric \cite{landsiedel2012low}. As described in \cite{landsiedel2012low}, the $EDC_i$ of node $i$ for a given subset $S_i$ of neighbors with link quality $p_{ij}$ and $EDC_j$ $(j\in Si)$ is as follows:
\begin{equation}\label{edc}
EDC_i(S_i) = \frac{1}{\sum\nolimits_{j\in S_i}p_{ij}} + \frac{\sum\nolimits_{j\in S_i}p_{ij}EDC_j}{\sum\nolimits_{j\in S_i}p_{ij}} + \omega
\end{equation}
The first term is the single hop EDC, which denotes how many units of time it requires on average to transmit a packet to one of the neighboring nodes in $S_i$. The second term describes the routing progress that the neighboring nodes in $S_i$ offer, weighted by their link qualities $p_{ij}$ (estimated from Packet Reception Rate (PRR)). The third term, $\omega$, adds a weight to reflect the cost of forwarding.

With a DODAG topology, the root node has an EDC of $0$. We select node $6$ from the experimental measurement used in our work as the root node because of its suitable position (with better proximity) and communication with all other nodes, hence $EDC_{node6} = 0$. The forwarding cost $\omega$ is chosen to be $0.1$ according to \cite{landsiedel2012low} as a good balance between energy efficiency, delay and reliability. EDC selects the forwarder sets (the nodes that can forward/relay the data) for each node from the neighbors. According to ORPL, node $a$ will only forward to a receiver node $b$, if and only if it has the destination on its routing set and,
\begin{equation*}\label{fwd}
\begin{multlined}
\hspace{1.8em} (EDC_b + \omega < EDC_a) \enspace \cap \enspace (p_{ab} > 50\%)\\ 
\hspace{9em} \textrm{(when routing upwards towards root)}\\\\
\hspace{-5em} (EDC_a < EDC_b + \omega) \enspace \cap \enspace (p_{ab} > 50\%)\\ 
\hspace{4em} \textrm{(when routing downwards towards destination)}
\end{multlined}
\end{equation*}
If there are multiple successful forwarders the node will select the forwarder with the best EDC or, link quality. ORPL uses randomized periodic broadcast for updating the routing information with a Trickle timer (based on Trickle algorithm \cite{levis2008emergence,levis2004trickle}). Here, for the wakeup interval of the trickle timer, the lower bound is set to be $400$ ms and the upper bound is set to be $1000$ ms. We analyzed the performance of ORPL with different redundancy constants (e.g., $k=2,3,4$) of the trickle timer and we obtained nearly identical results. We chose the redundancy constant as $k = 4$ according to \cite{duquennoy2013let,clausen2012experiences}, where $k = 3$ to $5$ is investigated as an optimal redundancy constant in deployments.%

\paragraph{LOADng} We implement LOADng based on a distance-based routing followed from traditional AODV protocol \cite{perkins2003ad}. Here, the RSSI values from the measurements are used for anticipating the distance between particular nodes and estimating the hop count to reach from source to destination. However, BAN radio propagation is dominated by local variations and not by distance-based path losses \cite{hanlen2009interference}. As a reactive protocol, LOADng routes are updated whenever a packet is lost due to a broken link condition. Besides being a reactive protocol, LOADng also uses a Route Hold Time (RHT) \cite{vuvcinic2013performance} to specify the lifetime of a route. After the RHT expires, the router updates the route to ensure the selection of a shortest path. Here, we have used a RHT of $500$ ms.

We consider outage probability, throughput (successful packets/s), average end-to-end delay and average energy consumption (per packet) as the performance metrics for evaluating the experimentally-based optimization techniques in this paper. Furthermore, we demonstrate the percentage of hop count induced from different protocols. Additionally, the statistical distribution fits for SPR and CMR are investigated.%

\subsection{Outage Probability}
\begin{figure}[!t]
\centering{\includegraphics[width=\figwidth]{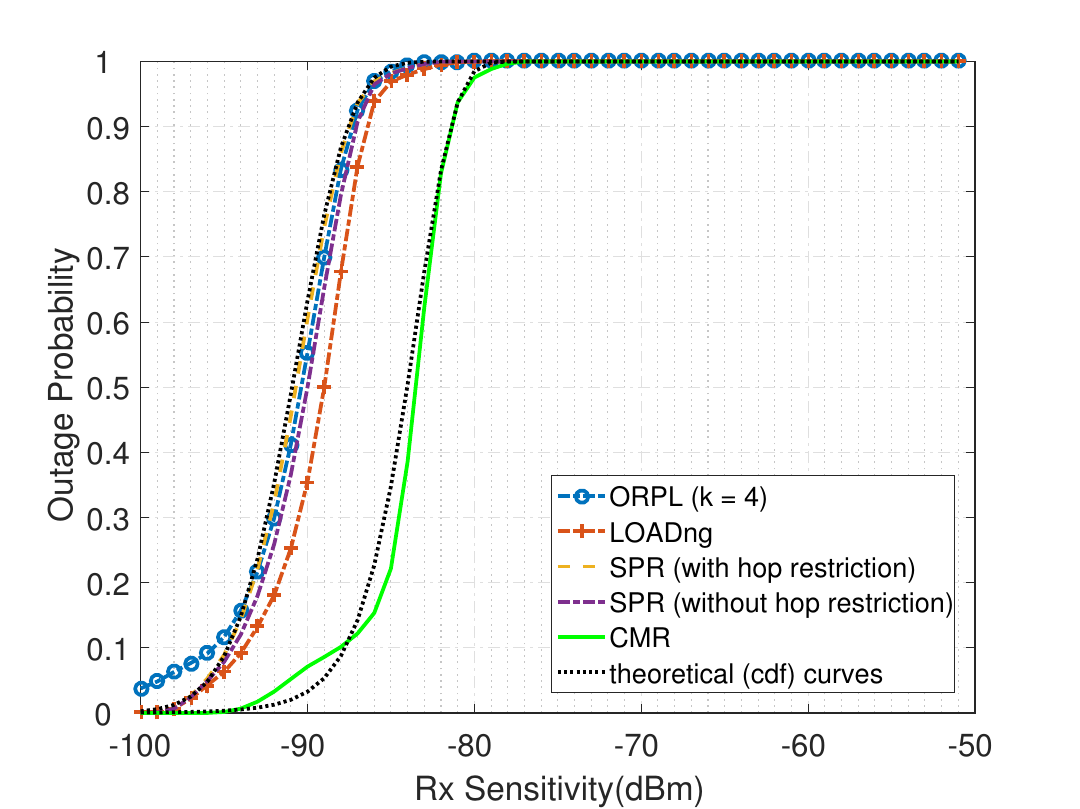}}
\caption{Outage probability of the averaged gains (over the network with $10$ BANs) found from ORPL (routing metric: EDC, k = $4$), LOADng (routing metric: hop count), SPR and CMR (routing metric: ETX, hop count); transmit power $0$ dBm. Black dotted curves represent the theoretical cdf (cumulative distribution function) of the corresponding outage probability and are well aligned.}
\label{outp1}
\end{figure}

An estimation of packet error rate and packet delivery ratio, and hence general performance can be made from outage probability, since it is the cumulative distribution function of channel gains. The average outage probability for a network of $10$ co-located mobile BANs with SPR and CMR, is presented in Fig. \ref{outp1} and can be expressed as follows:
\begin{equation}\label{outp_eqn}
 P_{out_x} = Prob\big(x_{comb} < rs_{th}\big)
\end{equation}
where, $P_{out_x}$ is the probability of combined channel gain $x_{comb}$ across routing technique $x$, being less than a given receive sensitivity threshold, $rs_{th}$.
The outage probabilities are taken from the overall network at assuming different receive sensitivities at a transmit power of $0$~dBm. For estimating the outages properly, the effect of non-recorded measurements (NaN) due to incorrectly decoded packets were replaced with a value of $-101$~dBm, just below the receiver sensitivity of $-100$~dBm.
In Fig. \ref{outp1}, it can be seen that SPR, LOADng and ORPL techniques have approximately similar results, where SPR (without hop restriction) has only marginally better packet delivery ratio than the SPR method with hop restriction, as it takes the path with lowest cost (where most of the routes consist of one or two hops) but can consume more energy and network lifetime due to the length of some routes (up to eight hops). LOADng provides slightly better packet delivery ratio than SPR because of the immediate route repair technique, however it can also take lengthy routes without hop count restriction. In that case, the routes chosen by SPR with respect to the combined metric (ETX + maximum two hops) provide a good trade-off between throughput and energy consumption (as shown in following subsections), as it is restricted to two hops. On the other hand, CMR (which uses a similar hop count restriction) provides up to $8$ dB, $7$ dB, and $6$ dB improvement over ORPL, SPR with hop restriction, and LOADng (as well as SPR without hop restriction), respectively, at $10\%$ outage probability. It is shown that in the best-case scenario (at $-100$ dBm receive sensitivity), there is $0$\% outage probability for $10$ hubs with all the protocols (with ORPL, that is $0.4\%$), which indicates a packet error rate close to 0\% ($< 10$\%), thus achieving the requirement of the IEEE $802.15.6$ Standard. Importantly, SPR (along with LOADng) and CMR techniques achieve less than 10\% outage probability at $-95$~dBm and $-88$~dBm receive sensitivity, respectively.
\begin{figure}[!t]
\centering{\includegraphics[width=\figwidth]{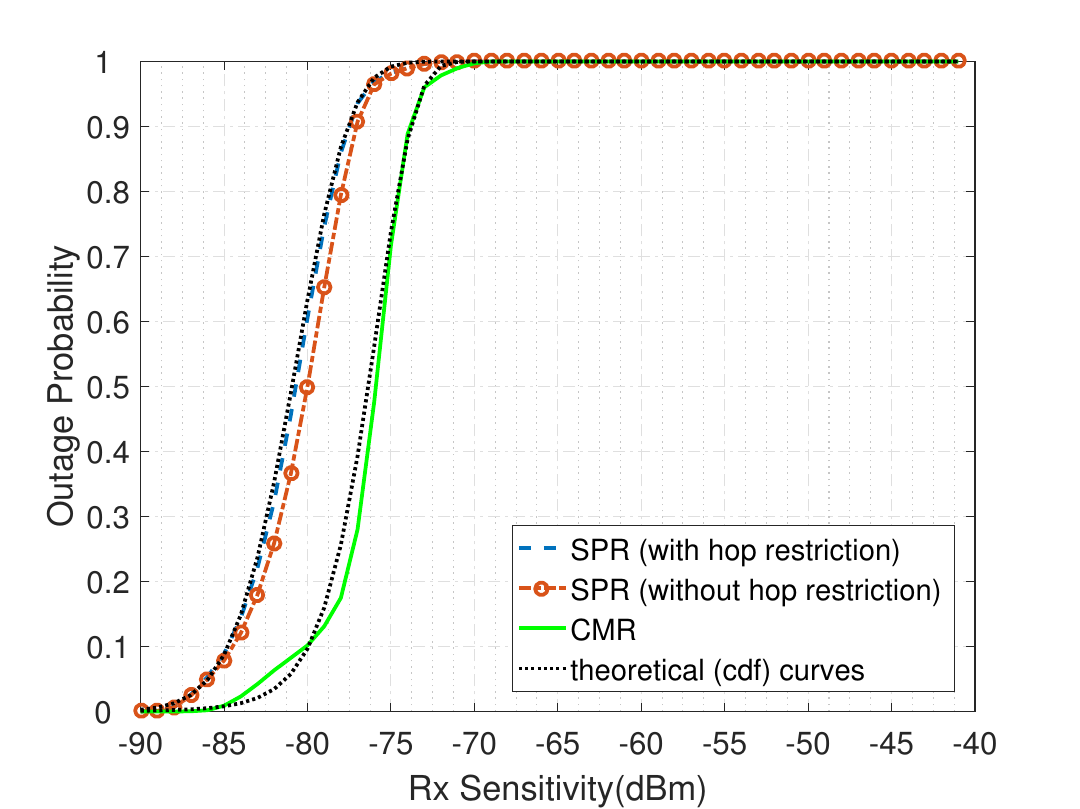}}
\caption{Outage probability of the averaged gains (over the network with $10$ BANs) found from SPR and CMR (routing metric: ETX, hop count); transmit power $10$ dBm (at hubs) and $5$ dBm (at relays). Black dotted curves represent the theoretical cdf (cumulative distribution function) of the corresponding outage probability and are well aligned.}
\label{outp2}
\end{figure}

The same process is repeated for SPR and CMR with transmit power $10$ dBm (at hubs) and $5$ dBm (at relays) in Fig. \ref{outp2}. The minimum receive sensitivity is considered to be $-90$ dBm, owing to the increased transmit power, which causes the lower limit of measured channel gains to be at $-100$~dB. For the same reason, the actual receiver sensitivity of the sensors is considered to be $-86$~dBm for calculating ETX values. By comparing Figs. \ref{outp1} and \ref{outp2}, it can be seen that the curves in Fig. \ref{outp2} shift right with the change of transmit power and receive sensitivity, which implies that with increased transmit power of on-body nodes, excellent reliability is obtained with less sensitive receivers. In this case, the average outage probability for $10$ hubs is also less than $10$\% for both SPR and CMR, for receive sensitivities of $\le$$-84$~dBm and $\le$$-80$~dBm, respectively.%

\subsection{Throughput (successful packets/s)}
\begin{figure}[!t]
\centering{\includegraphics[width=\figwidth]{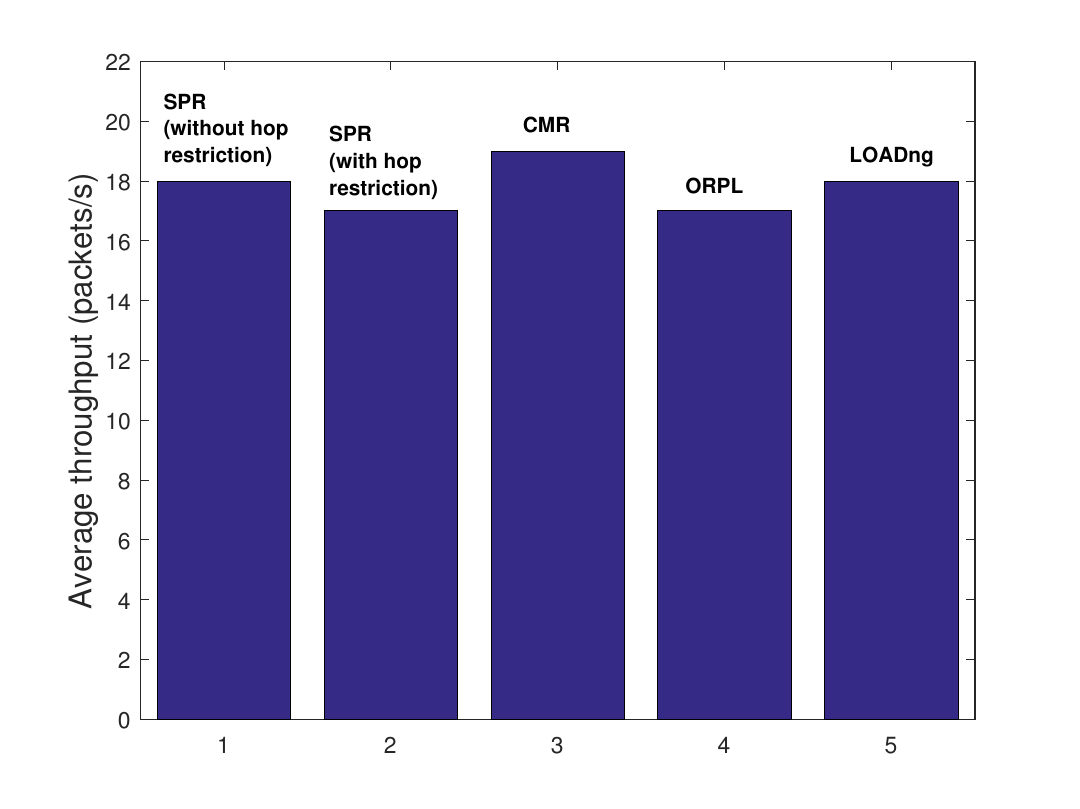}}
\caption{Average throughput (packets/s) for SPR, CMR, ORPL, and LOADng; at $-100$ dBm receive sensitivity with transmit power $0$ dBm.}
\label{thr}
\end{figure}

Throughput provides an assumption of how much information can be transferred per unit time (with a given transmission rate). The throughput (in packets/s) is analyzed considering that in each transmission period (or active period) the node will transmit at least one packet, however given the transmission period and packet transmission time (Table \ref{table_param}), more than one packet can be transmitted during the transmission period. We then estimate the throughput, $\Theta_x$ (number of successful packets per second) based on the received signal strength at the destination for different protocols as follows:
\begin{equation}
\Theta_x = \frac{\sum\big(x_{comb} \geq rs_{th}\big)}{Total\_time\text{ } (s)}
\end{equation}
where $\Theta_x$ is the total number of successful packets (estimated from the combined channel gain $x_{comb}$ across routing technique $x$, with respect to the given receive sensitivity threshold, $rs_{th}$) per unit time (second). The number of successful packets is averaged over the whole network at continuous times. The average throughput over the network with $10$ BANs with different protocols is shown in Fig. \ref{thr}.
It can be seen that CMR outperforms the other protocols in terms of throughput by providing $95\%$ successful packet delivery ($19$ packets/s at a packet transmission rate of $20$ Hz), which indicates increased successful transmissions with multi-path routing incorporating cooperative combining. Also, SPR (with hop restriction) and ORPL has lower throughput with respect to other protocols.%

\subsection{Delay}
Delay is an important design and performance characteristic of network. The term delay used throughout this paper is referred to as the end-to-end delay, which specifies how long it takes for a data packet to travel across the entire network path from source to destination. End-to-end delay can be roughly estimated as follows:
\begin{equation}\label{eq_delay}
 D_{end\_end} = D_{trans} + D_{queue} + D_{proc} + D_{prop},
\end{equation}
where $D_{trans}$, $D_{queue}$, $D_{proc}$, and $D_{prop}$ are the transmission, queuing, processing, and propagation delays, respectively. As the processing and propagation delays are negligible in the scenario described in this paper, we have calculated transmission and queuing delays to evaluate the end-to-end delay. The transmission delay ($D_{trans}$) is the delay for packet transmission ($T_{packet}$). The queuing delay is referred to as the waiting delay that occurs at intermediate nodes (hubs/relays) during a packet delivery. As each node is transmitting every $50$ ms without central coordination, we consider the highest possible amount of waiting period (e.g., $49$ ms, excluding the packet transmission time of around $1$ ms) when estimating the waiting/queuing delay at intermediate BANs/relays. In case of retransmission due to the failure of delivering a packet, the amount of $D_{end\_end}$ is doubled. In CMR, the on-body sensors/relays (in each route-hop) are operating in a different tier and coordinated by their hub with a sequential transmission. Again, we consider the highest amount of waiting delay for the intermediate BAN/relay node in CMR, e.g., $59$ ms considering the extra waiting time for the packet transmission of the cooperative relayed links at the route-hop. It should be noted that in practical situations the end-to-end delay may be less than the estimated delay here, e.g., the waiting period at an intermediate node can be lower than the maximum amount, as the packet can arrive at any time during the transmission period and the intermediate BAN hub can transmit at any time during the sampling period.

\begin{figure}[!t]
\centering{\includegraphics[width=\figwidth]{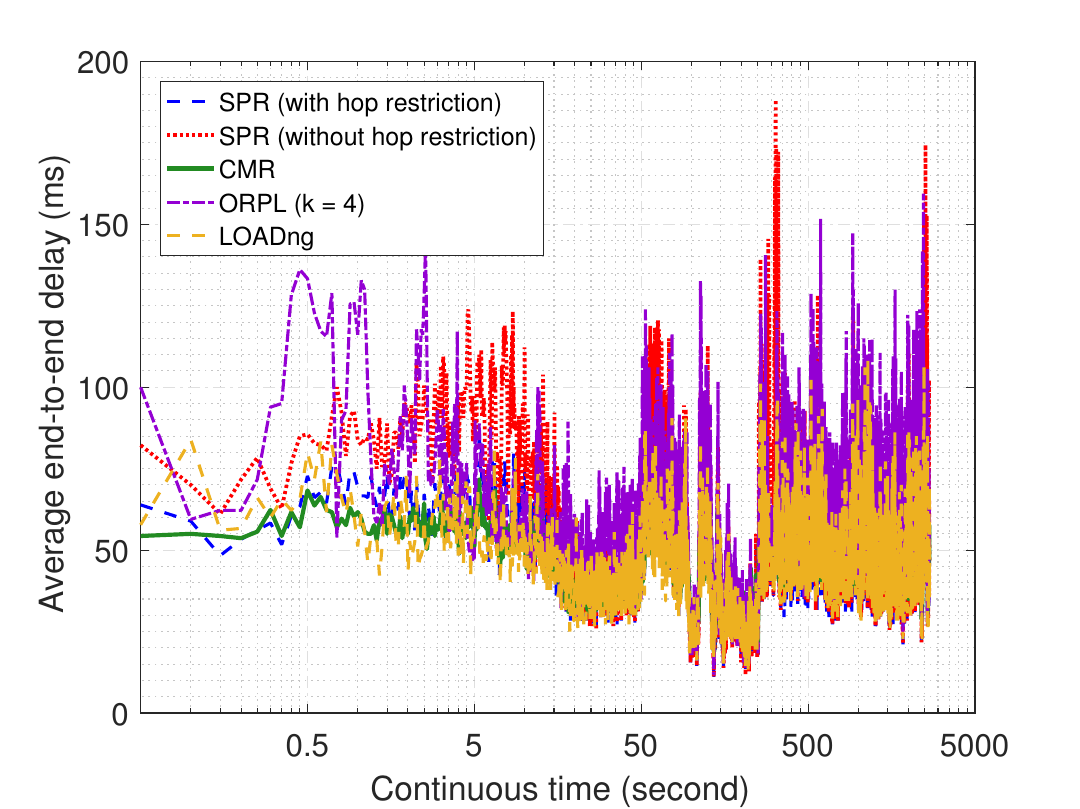}}
\caption{Average end-to-end delay at continuous times for SPR, CMR, ORPL, and LOADng; at $-100$ dBm receive sensitivity with transmit power $0$ dBm.}
\label{Del}
\end{figure}

\begin{figure}[!t]
\centering{\includegraphics[width=\figwidth]{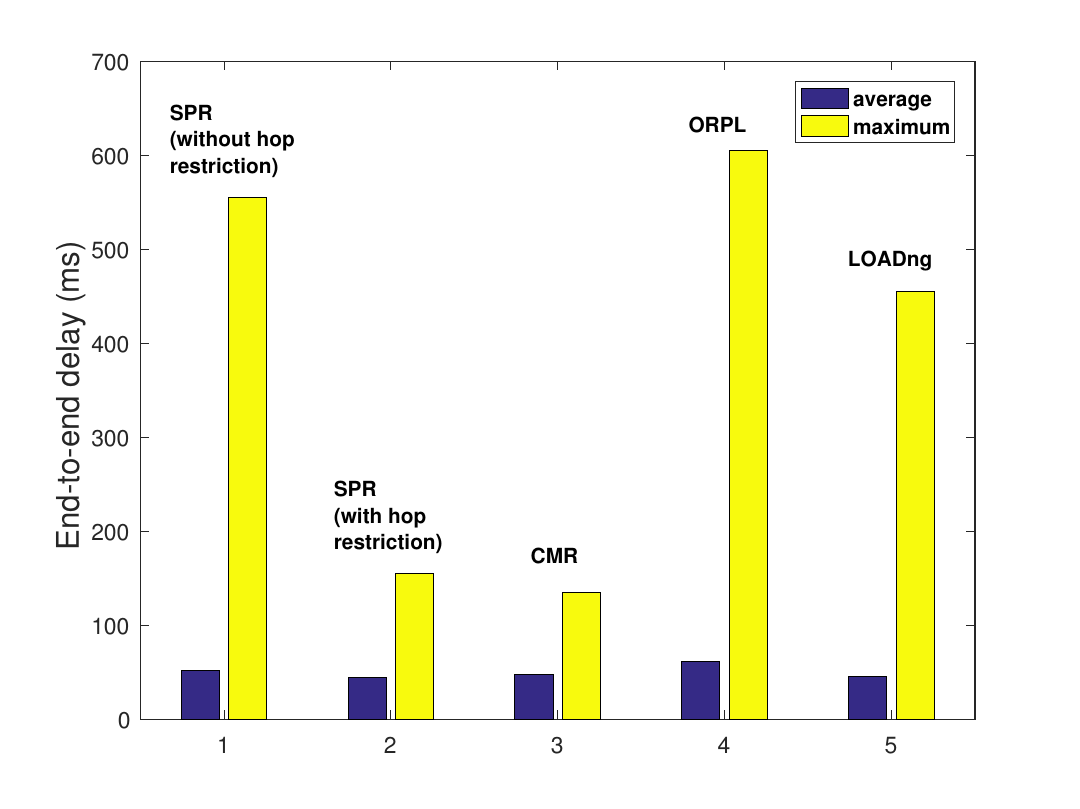}}
\caption{Average and maximum end-to-end delay over the whole period for SPR, CMR, ORPL, and LOADng with the network consisting of $10$ BANs; with transmit power $0$ dBm at $-100$ dBm receive sensitivity.}
\label{Del_bar}
\end{figure}

The average end-to-end delays at continuous times with different protocols over the whole network consisting of $10$ co-located BANs are presented in Fig. \ref{Del}. According to the IEEE $802.15.6$ Standard, latency should be less than $125$  ms in medical applications and less than $250$ ms in non-medical applications \cite{tg6_std}. The average and maximum end-to-end delay over the whole period of the BBN with different protocols are shown in Fig. \ref{Del_bar}, where all of them are producing an acceptable amount of delay (on average) according to the IEEE $802.15.6$ BAN Standard for medical and non-medical applications. As the average delay is estimated for the specific dataset with a finite amount of channel measurements, the knowledge of the maximum amount of delay (with the dataset) is beneficial in the long run to get an assumption of the upper limit of the delay that can be caused by the protocols. Although CMR generates slightly increased delay (still acceptable) on average than SPR (with hop restriction) and LOADng due to the selection combining at route-hops, the maximum delay caused by CMR is lower then all other protocols owing to the reduced retransmissions (yet producing the highest throughput). The maximum amount of delay of CMR and SPR (with hop restriction) is less than $250$ ms, which is an acceptable amount of latency for BAN non-medical applications. The maximum delay generated by other protocols are very high, specially for ORPL, which produces the highest amount of delay ($61$ ms on average, can go up to $605$ ms), as the delay increases when there is no possible forwarding node in the forwarding set. In all of the cases, the transmit power and receive sensitivity are $0$ dBm and $-100$ dBm, respectively, across all nodes (hubs/relays).

We also estimate the average end-to-end delays with differing receive sensitivities (e.g., $-90$ dBm, $-86$ dBm) for different protocols (results are shown in Table \ref{table_results}). It can be seen from Table \ref{table_results} that with respect to CMR, the average delays for other protocols (e.g., SPR, ORPL, LOADng) increase significantly, due to the added retransmission delays for increased packet failure rate and longer paths taken for reliable packet delivery with less receive sensitivity. More importantly, with less receive sensitivity (e.g., $-90$ dBm, $-86$ dBm) at the same transmit power, i.e., $0$ dBm, the average amount of delay for CMR remains lower than all other protocols.%

\subsection{Energy Consumption}
Energy consumption is one of the main performance metrics for resource-constrained networks like BANs. The energy consumption for each transmitted packet is calculated as follows:
\begin{equation}\label{eneg1}
E_p = \mathlarger{\sum}_{i=1}^{h}E_{packet_i} + \mathlarger{\sum}_{j=1}^{n}E_{idle_j}
\end{equation}
where $E_{packet_i}$ is the energy consumption for packet transmission in $i^{th}$ hop and $E_{idle_j}$ is the energy consumed by the $j^{th}$ transceiver in idle period during packet transmission. $h$ and $n$ are the number of hops and intermediate nodes/relays of a given route from source to destination, respectively.

$E_{packet}$ and $E_{idle}$ for each hop are calculated as follows:
\begin{equation}\label{eneg2}
E_{packet} =
\begin{cases}
T_{packet} \times (P_{TX_h} + P_{RX_h}), \quad \textrm{for hub-to-hub}\\
T_{packet} \times (P_{TX_h} + P_{RX_s}), \quad \textrm{for hub-to-sensor}\\
T_{packet} \times (P_{TX_s} + P_{RX_h}), \quad \textrm{for sensor-to-hub}
\end{cases}
\end{equation}%

\begin{equation}\label{eneg3}
E_{idle} = T_{idle} \times P_{idle}
\end{equation}
where $T_{packet}$ is the packet transmission time, $T_{idle}$ is the idle period of the transceiver during a packet transmission, $P_{TX_h}$/$P_{RX_h}$ are the power consumption of the on-body hubs during TX/RX mode, $P_{TX_s}$/$P_{RX_s}$ are the power consumption of the on-body sensors during TX/RX mode and $P_{idle}$ is the power consumption in idle mode. The applied parameters for the energy consumption estimation are given in Table \ref{table_param}.
\begin{figure}[!t]
\centering{\includegraphics[width=\figwidth]{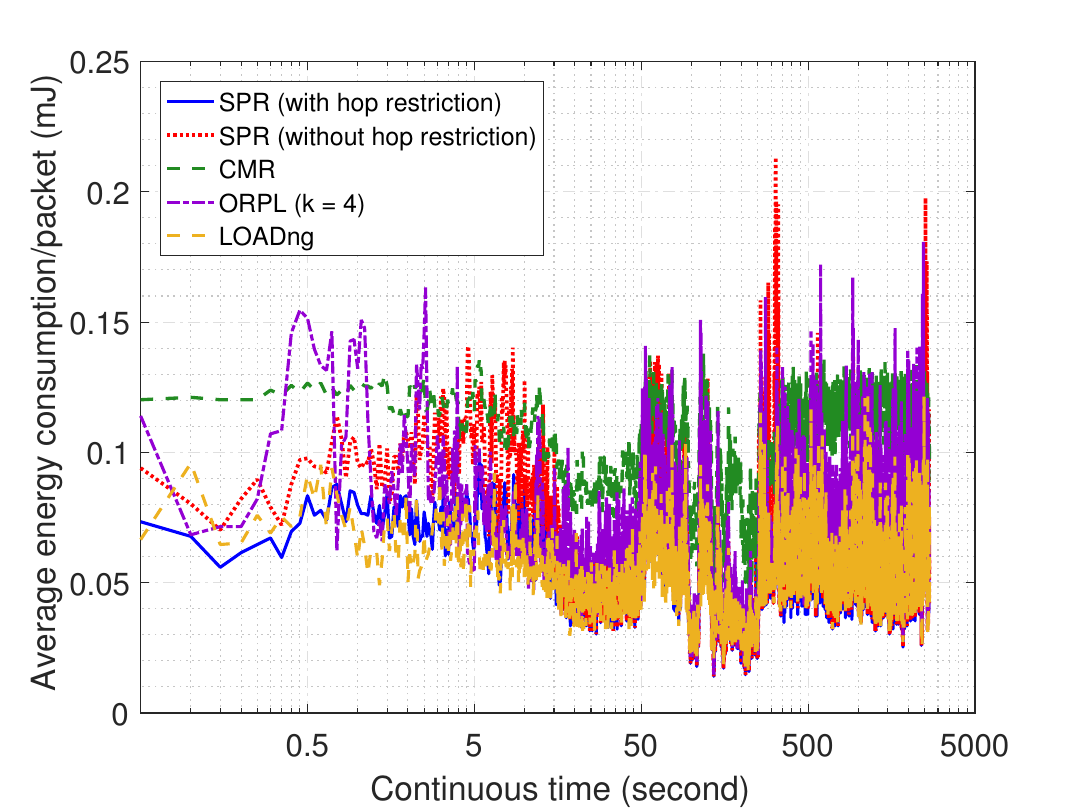}}
\caption{Average energy consumption (per packet delivery) at continuous times for SPR, CMR, ORPL, and LOADng; at $-100$ dBm receive sensitivity with transmit power $0$ dBm.}
\label{eneg_fig1}
\end{figure}
\begin{figure}[!t]
\centering{\includegraphics[width=\figwidth]{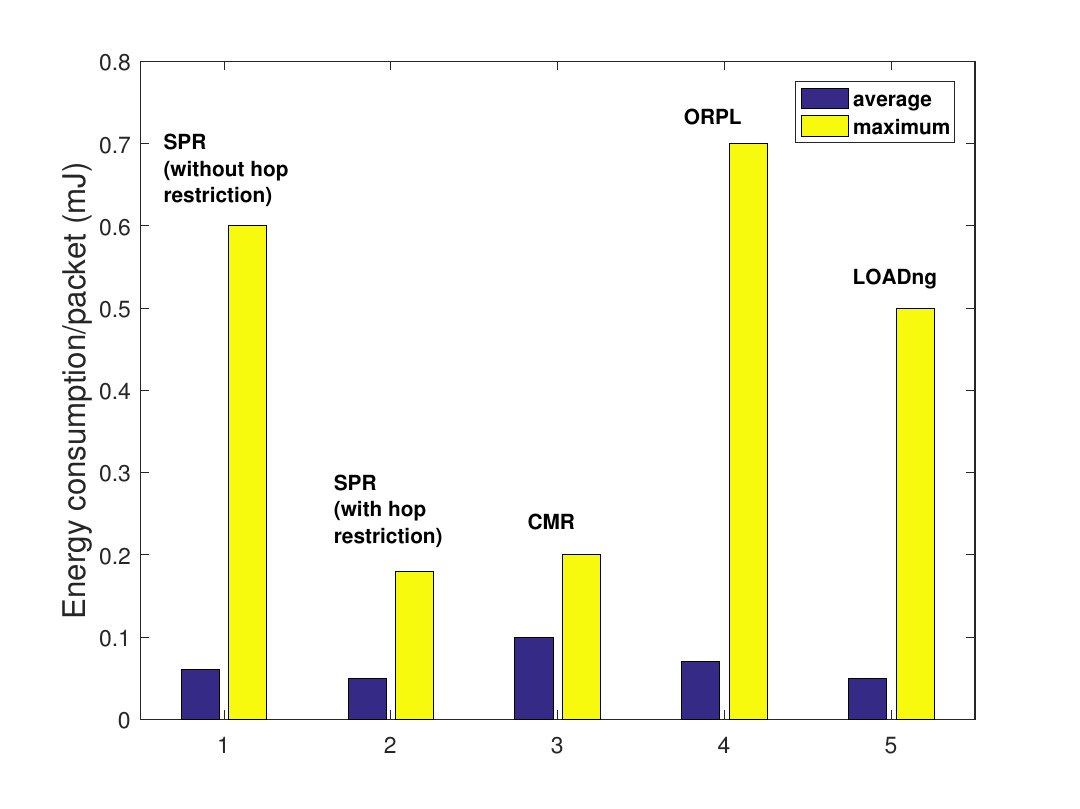}}
\caption{Average and maximum energy consumption (per packet delivery) over the whole period for SPR, CMR, ORPL, and LOADng with the network consisting of $10$ BANs; with transmit power $0$ dBm at $-100$ dBm receive sensitivity.}
\label{eneg_bar}
\end{figure}%

The average energy consumption (per packet delivery) at continuous times over the whole network with different protocols is shown in Fig. \ref{eneg_fig1}. Also, the average and maximum energy consumption of the network over the whole measurement period is shown in Fig. \ref{eneg_bar}. From Fig. \ref{eneg_bar}, reasonably, the average energy consumption over the total time for CMR ($0.1$ mJ) is more than the other protocols because of the extra energy consumption of the on-body sensors in each route hop. But CMR consumes less energy ($0.2$ mJ) with respect to other protocols (except SPR with hop restriction) in terms of maximum energy consumption, e.g., ORPL can consume up to $0.7$ mJ energy per packet delivery. Although LOADng consumes lower energy on average ($0.05$ mJ) because of its highly reactive characteristic, the energy consumption can rise $10$ times higher ($0.5$ mJ) when considering the maximum energy consumption, which can further increase in high-density networks with larger hop counts \cite{vuvcinic2013performance}. Also, it can be seen from Fig. \ref{eneg_fig1} that in some cases, the other protocols (e.g., SPR without hop restriction, ORPL, LOADng) consume more energy than CMR, owing to the extra energy consumption of an increased number of hops and from retransmission of packets in case of failure. Interestingly, from Table \ref{table_results}, when lowering the receive sensitivity (e.g., $-90$ dBm, $-86$ dBm) with the same transmit power, the average energy consumption increases for all the protocols except CMR. In this case, CMR in fact helps to improve energy consumption by reducing the packet failure rate and retransmissions with a cooperative path, having a possible route (up to two hops) with a lower idle period.

However, it is plausible to further reduce the energy consumption by optimizing the overhearing and broadcasting transmissions of nodes, reducing the overhead of frequent periodic updates of the routing table (updating when necessary) and opportunistic relaying in route hops of CMR. Besides that, if the hub can be carried by the BAN (e.g., as a smart phone or any other device) rather than worn as an on-body sensor, then the energy consumption will cause limited overhead for hub-to-hub communications.
\begin{figure*}[!t]
\centering
\subfloat[SPR (without hop restriction)]{\includegraphics[width=.3\textwidth]{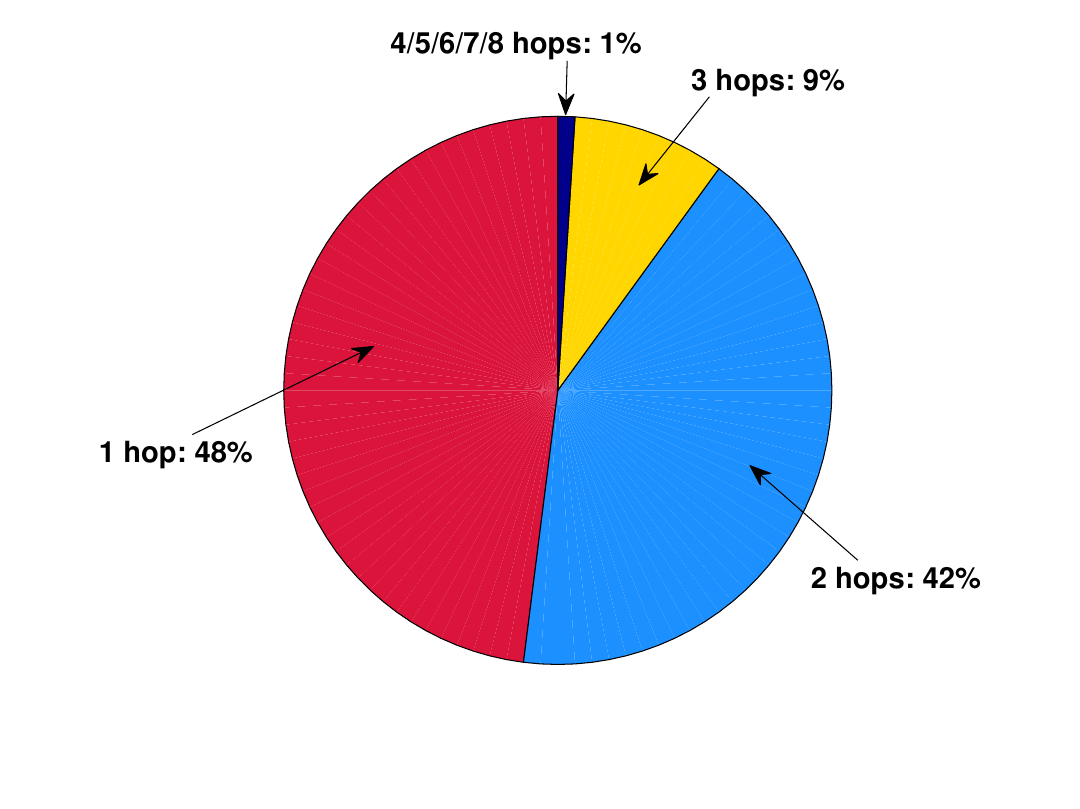}}
\hfill
\subfloat[ORPL (k = $4$)]{\includegraphics[width=.3\textwidth]{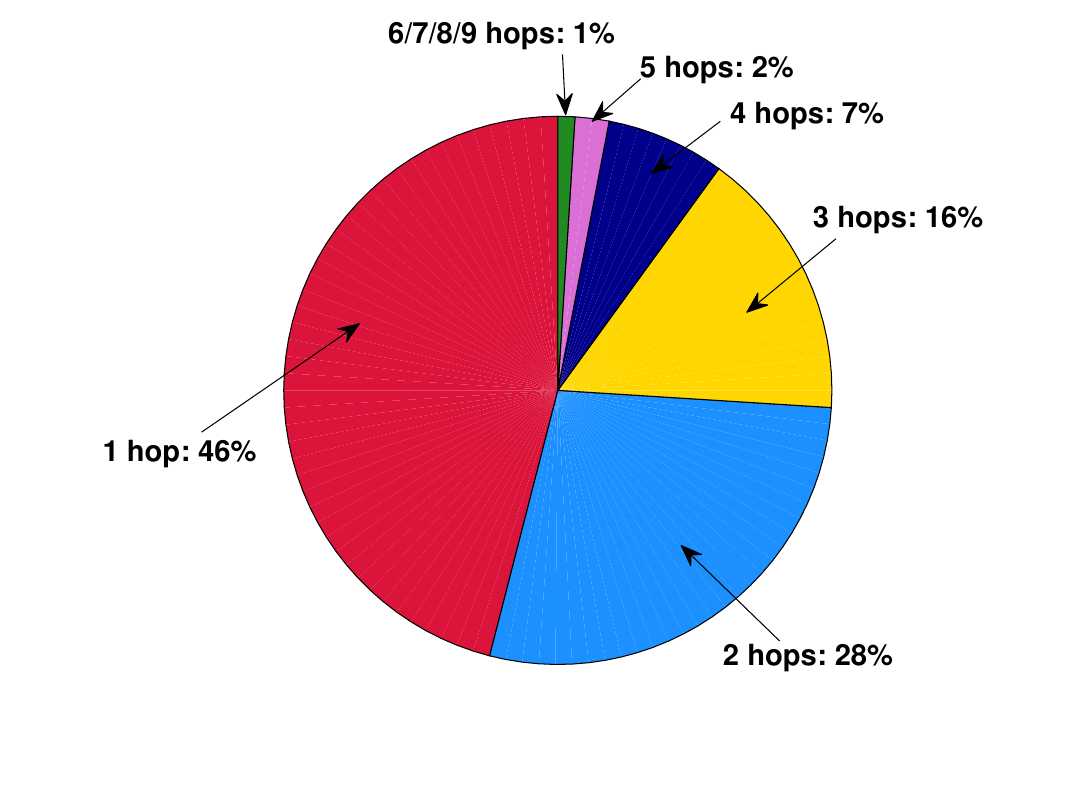}}
\hfill
\subfloat[LOADng (RHT = $500$ ms)]{\includegraphics[width=.3\textwidth]{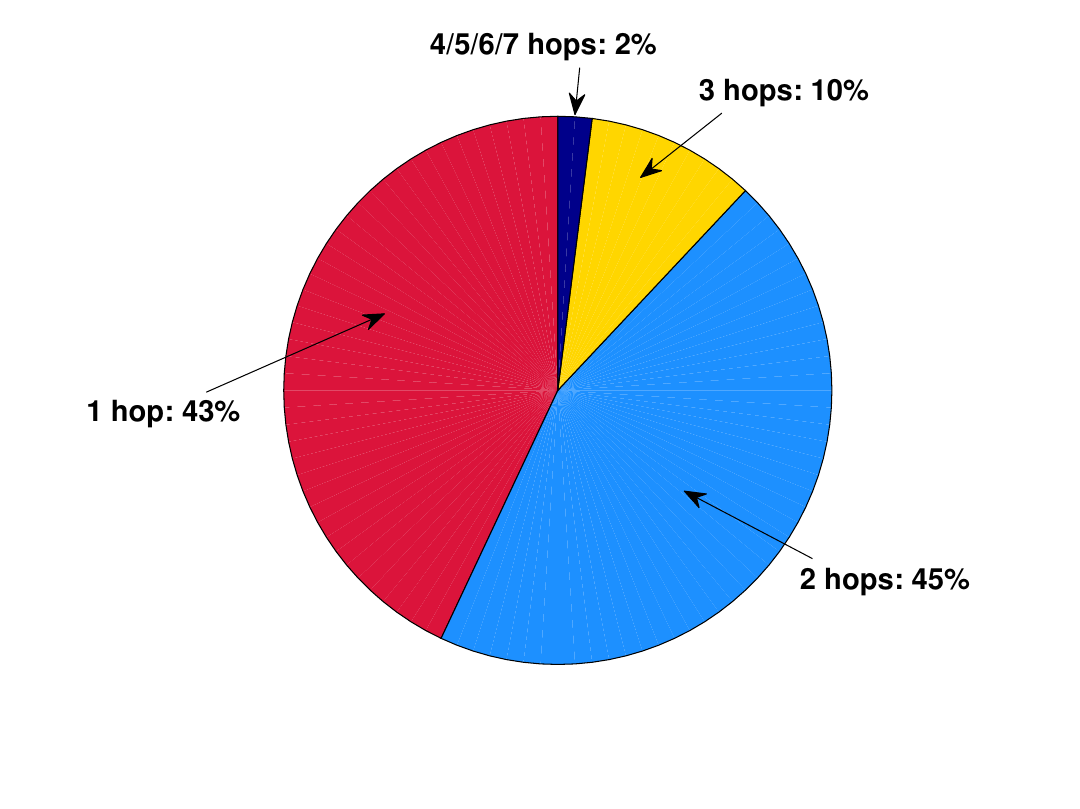}}
\caption{Percentage of hop count of routes with different protocols (SPR, ORPL, LOADng), at $-100$ dBm receiver sensitivity.}
\label{hop_count}
\end{figure*}%

\subsection{Percentage of Hop Count}
The hop count percentage of routes taken by a given protocol indicates how much overhead (e.g., delay, energy consumption) is caused by that protocol. We have investigated the percentage of different number of hop counts with different protocols (with $-100$ dBm receiver sensitivity), as shown in Fig. \ref{hop_count}. It can be seen in Fig. \ref{hop_count} that a major portion of the time for all protocols is occupied by routes that consist of one hop (around $45\%$ on average) or two hops (around $38\%$ on average). On the other hand, the ORPL protocol takes more than two hops (up to nine hops) for a significant amount of time ($26\%$ of the total period) because of the DODAG topology, which increases the overall delay and energy consumption due to extra overhead, e.g., finding forwarding node which has the destination on its routing set. Additionally, it is found that with the EDC metric (which partially depends on the quality of the links connected to the root node), there is no possible route for up to $13\%$ of the time with ORPL, as ORPL only takes the forwarder nodes which has PRR greater than $50\%$. Hence, the restriction to a maximum of two hops (in conjunction with the ETX metric) is a suitable choice to jointly optimize energy consumption and reliability, while incorporating less overhead. However, with a maximum two hop restriction, SPR and CMR choose direct (one hop) links more than $50\%$ of the time, as it chooses the best path (with one or two hops) that has a smaller transmission/retransmission count. Furthermore, as shown in this section, we have gained an acceptable outage probability and delay performance (compliant with the IEEE $802.15.6$ Standard) by the use of the combined metric (ETX + maximum two hops).

The empirical results found from Figs. \ref{outp1} to \ref{eneg_bar}, are summarized in Table \ref{table_results}.
\begin{table*}[!t]
\centering
\caption{Empirical result analysis for different protocols (e.g., SPR, CMR, ORPL, LOADng)}
\label{table_results}
\renewcommand{\arraystretch}{1.2}
\begin{tabular}{|l|c|c|c|c|c|}\Xhline{1pt}
 & \begin{tabular}{@{}c@{}c@{}} \textbf{SPR}\\\textbf{(without}\\\textbf{hop restr-}\\\textbf{iction)}\\[0.05cm]\end{tabular} & \begin{tabular}{@{}c@{}c@{}} \textbf{SPR}\\\textbf{(with hop}\\\textbf{restriction)}\\[0.05cm]\end{tabular} & \begin{tabular}{@{}c@{}} \textbf{CMR}\end{tabular} & \begin{tabular}{@{}c@{}} \textbf{ORPL}\\\textbf{(k = 4)}\\[0.05cm]\end{tabular}
& \begin{tabular}{@{}c@{}} \textbf{LOADng}\\\textbf{(RHT =}\\\textbf{$500$ ms)}\\[0.05cm]\end{tabular}\\\Xhline{0.8pt}
\begin{tabular}{@{}l@{}l@{}l@{}}Best-case outage probability\\(at $-100$ dBm receive sensitivity)\\with transmit power $0$ dBm\\[0.05cm]\end{tabular} & $0$\% & $0$\% & $0$\% & $0.4$\% & $0$\%\\\hline
\begin{tabular}{@{}l@{}l@{}l@{}l@{}}Performance improvement\\of CMR over other protocols\\at $10$\% outage probability\\with transmit power $0$ dBm\\[0.05cm]\end{tabular} & $6$ dB & $7$ dB & - & $8$ dB & $6$ dB\\\hline
\begin{tabular}{@{}l@{}l@{}}Average throughput (packets/s)\\at $-100$ dBm receive sensitivity\\[0.05cm]\end{tabular} & $18$ & $17$ & $19$ & $17$ & $18$\\\Xhline{0.8pt}
\begin{tabular}{@{}l@{}l@{}}Average end-to-end delay at\\$-100$ dBm receive sensitivity\\[0.05cm]\end{tabular} & $51.7$ ms & $44.3$ ms & $47.5$ ms & $61$ ms & $45.9$ ms\\\hline
\begin{tabular}{@{}l@{}l@{}}Maximum end-to-end delay at\\$-100$ dBm receive sensitivity\\[0.05cm]\end{tabular} & $555$ ms & $155$ ms & $135$ ms & $605$ ms & $455$ ms\\\hline
\begin{tabular}{@{}l@{}l@{}}Average end-to-end delay at\\$-90$ dBm receive sensitivity\\[0.05cm]\end{tabular} & $61.9$ ms & $53.3$ ms & $50.5$ ms & $72.2$ ms & $57$ ms\\\hline
\begin{tabular}{@{}l@{}l@{}}Average end-to-end delay at\\$-86$ dBm receive sensitivity\\[0.05cm]\end{tabular} & $84.5$ ms & $72.3$ ms & $60$ ms & $97.1$ ms & $81.4$ ms\\\Xhline{0.8pt}
\begin{tabular}{@{}l@{}l@{}}Average energy consumption at\\$-100$ dBm receive sensitivity\\[0.05cm]\end{tabular} & $0.06$ mJ & $0.05$ mJ & $0.1$ mJ & $0.07$ mJ & $0.05$ mJ\\\hline
\begin{tabular}{@{}l@{}l@{}}Maximum energy consumption\\at $-100$ dBm receive sensitivity\\[0.05cm]\end{tabular} & $0.6$ mJ & $0.17$ mJ & $0.2$ mJ & $0.7$ mJ & $0.5$ mJ\\\hline
\begin{tabular}{@{}l@{}l@{}}Average energy consumption at\\$-90$ dBm receive sensitivity\\[0.05cm]\end{tabular} & $0.07$ mJ & $0.06$ mJ & $0.1$ mJ & $0.08$ mJ & $0.07$ mJ\\\hline
\begin{tabular}{@{}l@{}l@{}}Average energy consumption at\\$-86$ dBm receive sensitivity\\[0.05cm]\end{tabular} & $0.1$ mJ & $0.08$ mJ & $0.12$ mJ & $0.11$ mJ & $0.09$ mJ\\\Xhline{1pt}
\end{tabular}
\end{table*}%

\subsection{Probability Density Functions}
\begin{figure}[!t]
\centering{\includegraphics[width=\figwidth]{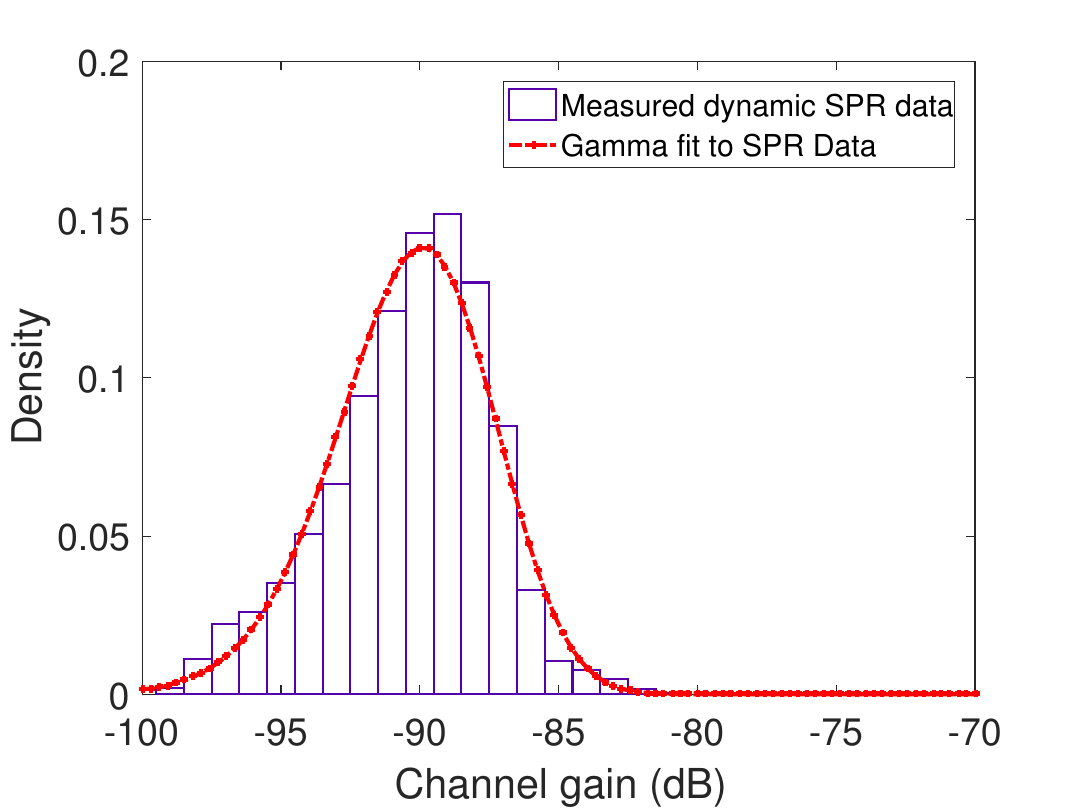}}
\caption{The empirical probability density of combined channel gain data from SPR with a gamma distribution fit, where $\kappa = 9.58$ and $\theta = 0.00000334$ are the shape and scale parameter, respectively.}
\label{gamfit}
\end{figure}

\begin{figure}[!t]
\centering{\includegraphics[width=\figwidth]{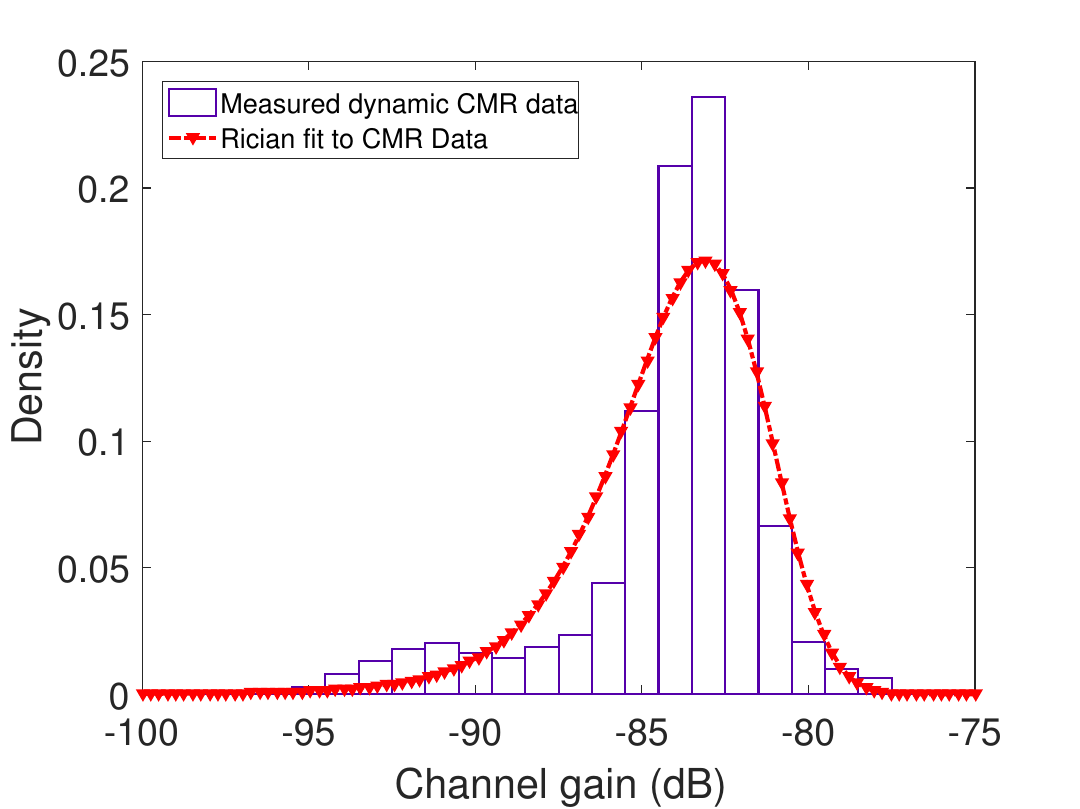}}
\caption{The empirical probability density of combined channel gain data from CMR with a Rician distribution fit, where $\nu = 0.0000626$ and $\sigma = 0.0000185$ are the two shape parameters.}
\label{ricefit}
\end{figure}

We investigate distribution fits, using typical statistical distributions, for the combined channel gains, acquired after SPR and CMR techniques are applied on the experimentally measured channel gain data. In Figs. \ref{gamfit} and \ref{ricefit}, it is shown that the probability density function of the combined channel gains across dynamic SPR and dynamic CMR provide gamma (Fig. \ref{gamfit}) and Rician (Fig. \ref{ricefit}) distribution fits, respectively. The maximum likelihood estimation (MLE) parameter is used to select the best fit. Channel gains are taken over $5329$ continuous timestamps, each of which is for $500$ ms.

The probability distribution for the combined channel after applying shortest path routing (SPR) over coexisting BANs is gamma, which can be calculated from
\begin{equation}\label{gamma}
f(x \mid \kappa,\theta) = \frac{1}{\theta^\kappa\Gamma(\kappa)}x^{\kappa - 1}\exp\left(\frac{-x}{\theta}\right), \quad x>0
\end{equation}
where $\kappa = 9.58$ and $\theta=0.00000334$ are the shape and scale parameter, respectively, for this channel gain fit after SPR is performed.

A Rice or Rician distribution (also known as a Nakagami-n distribution) models Rician fading, in which signal cancellations affect radio propagation \cite{gudbjartsson1995rician}. Rician fading occurs when the signal arrives at the receiver by several different paths and one of the paths (typically a line-of-sight path) is stronger than the others. Here, the probability distribution for the combined channel gains from cooperative multi-path routing (CMR) with coexisting BANs can be approximated from the Rician probability distribution as follows:
\begin{equation}\label{rician}
f(x \mid \nu,\sigma) = \frac{x}{\sigma^2}\exp\left(\frac{-(x^2 + \nu^2)}{2\sigma^2}\right)I_o\left(\frac{x\nu}{\sigma^2}\right)
\end{equation}
where $I_o(z)$ is the modified Bessel function of the first kind with order zero, and $\nu = 0.0000626$ and $\sigma = 0.0000185$ are the two shape parameters, for the combined channel gain fit after CMR is performed.%

\section{Conclusion}
In this paper, we have analyzed cross-layer methods, i.e., shortest path routing (SPR) and cooperative multi-path routing (CMR) with experimental measurements, to optimize radio communications across distributed wireless body-to-body networks (BBNs), by utilizing distinct features at the physical and network layers. Physical layer information (e.g., ETX, hop count) is dynamically fed into the network layer for determining real-time and reliable routes among BAN hubs. We have compared the performance of SPR and CMR with some state-of-the-art WSN protocols (e.g., ORPL, LOADng) that support any-to-any routing. From our analysis, it is evident that the ETX or the hop count metric with mesh topology can perform better than the EDC metric (with DODAG topology) in case of any-to-any routing, as ORPL performance (which makes use of the EDC metric) falls behind the other protocols. This also implies that SPR and CMR provides improvement over the CTP and RPL protocols (other state-of-the-art WSN protocols) as ORPL outperforms those protocols. 

We have shown that in the best-case scenario (at $-100$ dBm receive sensitivity), shortest path routing (SPR) and cooperative multi-path routing (CMR) along with other protocols provide negligible packet error rate and an acceptable amount of end-to-end delay (on average) according to the IEEE $802.15.6$ BAN Standard. At $10\%$ outage probability, CMR gives significantly better performance than other protocols by contributing up to $8$ dB, $7$ dB, and $6$ dB improvement over ORPL, SPR, and LOADng, respectively, occurring at practical receive sensitivities. The maximum amount of end-to-end delay with CMR ($135$ ms) is lowest amongst all protocols (e.g., ORPL generates a maximum end-to-end delay of $605$ ms) due to the reduced retransmissions and hop count restrictions. Notably, CMR outperforms other protocols in terms of throughput (successful packets/s) while providing acceptable amount (for medical/non-medical applications) of average end-to-end delay ($47.5$ ms), at $-100$ dBm receive sensitivity. Also, CMR provides the lowest amount of average end-to-end delay with respect to other protocols for reduced receive sensitivity (e.g., $-86$ dBm, $-90$ dBm). Our analysis shows that CMR consumes more energy (on average) than other protocols because of the power consumption in the cooperative relayed links at route-hops. However, the maximum energy consumption with CMR is much lower than other protocols (excepting SPR with hop restriction). Also, in some cases, CMR in fact reduces the energy consumption by increasing packet success rate with less retransmissions. For instance, with less receive sensitivity (e.g., $-86$ dBm, $-90$ dBm), the average energy consumption for other protocols (e.g., SPR, ORPL, LOADng) increases while the average energy consumption CMR remains approximately the same.

Furthermore, it is demonstrated that most of the routes for all the protocols consist of one or two hops, thus validating the applicability of two hop restriction with the ETX metric to optimize the performance of BBN communications in real life scenarios. We have also observed that the combined channel gains across a complete SPR route with narrowband communications possess a gamma distribution, where the complete combined channel gains from CMR have a Rician distribution. This work provides feasible methods for the deployment of many closely-located BANs in decentralized real-world applications with large scale and highly connected medical/non-medical systems, and will be investigated further for more than $10$ BANs/people in future research for obtaining further understanding of body-to-body network (BBN) performance in other practical scenarios.%

\bibliographystyle{IEEEtran}
\bibliography{Reference}

\end{document}